\documentclass[twocolumn,english,letterpaper,showpacs]{revtex4}
\usepackage{babel,amsmath,amssymb}
\newtheorem{proposition}{Proposition}

\begin{document}

\title{Quasinormal modes of d-dimensional spherical black holes
with \\ a near extreme cosmological constant}

\author{C. Molina}

\email{cmolina@fma.if.usp.br}

\affiliation{Instituto de F\'{\i}sica, Universidade de S\~{a}o Paulo
\\ C.P. 66318, CEP 05315, S\~{a}o Paulo-SP, Brazil}


\begin{abstract}
We derive an expression for the quasinormal modes of scalar
perturbations in near extreme \mbox{$d$-dimensional} Schwarzschild-de
Sitter and Reissner-Nordstr\"{o}m-de Sitter black holes. We show that,
in the near extreme limit, the dynamics of the scalar field  is
characterized by a P\"{o}schl-Teller effective potential. The results
are qualitatively independent of the spacetime dimension and field mass.
\end{abstract}

\maketitle

\section{Introduction}

Recent observational results suggest that the universe in large scale
is described by an Einstein equation with an (at least effective)
cosmological constant. In this context, the dynamics of fields in
spacetimes which are not asymptotically flat has taken on new importance.
Quasinormal modes are a fundamental part of this dynamics. In general,
perturbations in the spacetime exterior of a black hole are followed
by oscillations which decay exponentially in time. Roughly speaking,
these are  quasinormal modes, complex frequency modes which carry information
about the background geometry, and are independent of the initial perturbation.
The characterization of the quasinormal modes originated from compact
objects are of particular relevance in gravitational wave astronomy 
\cite{Kokkotas-99}. Their detection is expected to be realized through
gravitational wave observations in the near future.

In asymptotically anti-de Sitter spacetimes, the anti-de Sitter/conformal
field theory (AdS/CFT) correspondence \cite{AdS-CFT} plays a very
important role. In this framework, a link is established  among
the quasinormal frequencies of a test field in AdS black holes and
the decay rates in the dual field theory. The first study of the
scalar quasinormal modes in AdS space was performed by Chan and Mann
\cite{Chan-97}. For Ba\~{n}ados-Teitelboim-Zanelli (BTZ) black holes,
exact expressions for the quasinormal modes were obtained by Cardoso
and Lemos for scalar, electromagnetic, and Weyl perturbations
\cite{Cardoso-01-1}. In this context, the connection of CFT and
quasinormal modes was explored by Birmingham \emph{et al.}
\cite{Birmingham}. The problem in higher dimensions was considered by
Horowitz and Hubeny \cite{Horowitz-00}. Their basic results were
confirmed in \cite{Wang-01} by direct calculation of the wave
functions. Some other perturbations in AdS backgrounds were treated in
\cite{perturb-AdS}.      

For asymptotically de Sitter spacetimes, similar conjectures have
been formulated. Strominger \cite{Strominger-01} proposed a holographic
duality relating quantum gravity on $d$-dimensional de Sitter
space (dS$_{d}$) to a conformal field theory residing on the past
boundary of dS$_{d}$. This dS/CFT correspondence has motivated several
works  (as, e.g., \cite{dS-CFT}). 

A very different and intriguing application of quasinormal modes have
been suggested recently. In the context of loop quantum gravity, it
has been shown that the area  of the event horizon is
quantized, but the expression involves a free quantity, the Barbero-Immirzi
parameter. Hod noticed that the asymptotic real part of the
quasinormal frequency of the Schwarzschild black hole is proportional
to $\ln 3$ \cite{Hod-98}. This observation was applied by Dreyer to fix
the  Barbero-Immirzi parameter \cite{Dreyer-03}. And in \cite{Motl-02}, Motl
proved Hod's observation. Kunstatter used these ideas 
\cite{Kunstatter-02} to propose a Bekenstein-Hawking entropy spectrum
for $d$-dimensional spherically symmetric black holes. In
\cite{Abdalla-03}, Abdalla, Castello-Branco, and Lima-Santos proposed
an area quantization prescription for the near extreme
Schwarzschild-de Sitter and Kerr black holes.  

Calculating analytic expressions for the quasinormal frequencies is usually
difficult, except in particular situations. Approximation schemes were
developed to circumvent this problem \cite{approx}. One case where
analytic expressions were found is the P\"{o}schl-Teller potential
\cite{Poschl-33}. For this potential, many properties have been proved
and the frequencies have been calculated
\cite{Ferrari-84,Beyer-99}. In a recent work \cite{Cardoso-03},
Cardoso and Lemos studied  the Schwarzschild-de Sitter geometry in
four dimensions, and showed that the dynamics is specified by a
P\"{o}schl-Teller effective potential. They calculated exact
expressions for the quasinormal modes, demonstrating why a previous
approximation made by Moss and Norman \cite{Moss-02} hold in the near
extreme regime.    

In the present work, we generalize the method used by Cardoso and
Lemos in \cite{Cardoso-03}. We analyze scalar fields, massless and
massive, in the $d$-dimensional Schwarzschild-de Sitter (SdS) and
Reissner-Nordstr\"{o}m-de Sitter (RNdS) geometry.  
The approach here is a bottom-up one. In the next section, the horizon
structure of the $d$-dimensional Schwarzschild-de Sitter is discussed. The
geometry of the block outside the event horizon is then treated in the
near extreme limit. In Sec. IV, we show that the ideas developed in
the previous sections can be applied to more general spacetimes, including the
$d$-dimensional Reissner-Nordstr\"{o}m-de Sitter black holes. In
Sec. V, the dynamics of the scalar field in spherical  backgrounds
is discussed, and   analytical formulas for the quasinormal
frequencies are calculated. In Sec. VI some final remarks are made.

\section{Schwarzschild-de Sitter Metric}

The metric describing a $d$-dimensional nonasymptotically flat
spherical black hole was   presented in 
\cite{Thangherlini-63}. Written in spherical coordinates, the
Schwarzschild-de Sitter metric is given by 
\begin{equation}
ds^{2}=-h(r)dt^{2}+h(r)^{-1}dr^{2}+r^{2}d\Omega_{d-2}^{2} \ ,
\label{metric}
\end{equation}
where the function $h(r)$ is
\begin{equation}
h(r)=1-\frac{2m}{r^{d-3}}-\frac{\Lambda r^{2}}{3} \ .
\end{equation}
The integration constant $m$ is proportional to the black hole mass, and
$d\Omega_{d-2}^{2}$ is the line element of the
($d-2$)-dimensional unit sphere:
\begin{eqnarray}
d\Omega_{d-2}^{2} & = & \left(d\theta^{1}\right)^{2}+\sin^{2}\theta^{1}\,
\left(d\theta^{1}\right)^{2} +\cdots \nonumber \\
& & +\sin^{2}\theta^{1}\cdots\,\sin^{2}\theta^{d-2}\,
\left(d\theta^{d-2}\right)^{2} \ .
\end{eqnarray}
If the cosmological constant is positive, the spacetime is asymptotically
de Sitter. In this case, $\Lambda$ is usually written in terms of
a ``cosmological radius'' $a$ as
\begin{equation}
\Lambda=\frac{3}{a^{2}} \ .
\end{equation}

The causal structure of the spacetime described by the metric presented
depends of the positive real roots of the function $h(r)$.
By rescaling the metric and the coordinates $t$ and $r$,
it can be seen that only one parameter is necessary  to characterize
the set of possible zeros of $h(r)$. Indeed, if the mass is
nonvanishing, a convenient parameter is $m^{2}/a^{2(d-3)}$.  We will
classify the possible horizons of the spacetime in the following
proposition. 

\begin{proposition} 
Let $d\geq4$, $a^{2}>0$ and $m>0$.
\begin{itemize}
\item The spacetime has two horizons if and only if the condition 
\begin{equation}
\frac{m^{2}}{a^{2(d-3)}} < \frac{(d-3)^{d-3}}{(d-1)^{d-1}}
\end{equation}
is satisfied. This is the usual $d$-dimensional Schwarzschild-de Sitter
black hole. 
\item The  spacetime has one horizon  if and only if the condition 
\begin{equation}
\frac{m^{2}}{a^{2(d-3)}} = \frac{(d-3)^{d-3}}{(d-1)^{d-1}}
\label{extreme_condition}
\end{equation}
is satisfied. This is the extreme Schwarzschild-de Sitter black
hole.
\item The  spacetime has no horizon if and only if the condition 
\begin{equation}
\frac{m^{2}}{a^{2(d-3)}} > \frac{(d-3)^{d-3}}{(d-1)^{d-1}}
\end{equation}
is satisfied. 
\end{itemize}
\end{proposition}

From proposition 1 we learn that the global structure of
the manifolds described by the Schwarzschild-de Sitter metric is
largely independent of the dimension. With a small value of
$m^{2}/a^{2(d-3)}$, this metric describes a spacetime with two
horizons --- an event horizon $r_{+}$ and a cosmological horizon
$r_{c}$ ($0<r_{+}<r_{c}$). The region of interest in this paper is the
block   
\begin{equation}
T_{+}=\left\{
(t,r,\theta_{1},\cdots,\theta_{d-2}),r_{+}<r<r_{c}\right\} \ .
\end{equation}
In the critical value, the event horizon and the cosmological horizon
coincide. This is the extreme Schwarzschild-de Sitter black hole.
When $m^{2}/a^{2(d-3)}$ is larger than the critical value, the metric
no longer describes a black hole.

\section{Near Extreme Schwarzschild-de Sitter  Black Hole}

Since we are interested in the limit where the event and cosmological
horizons are very close, it is natural to define the dimensionless
parameter $\delta'$ as
\begin{equation}
\delta'=\frac{r_{c}-r_{+}}{a} \ ,
\end{equation}
so that the near extreme limit is such that
\begin{equation}
0<\delta'\ll1\label{limit1} \ .
\end{equation}

In the Schwarzschild-de Sitter (SdS) black hole case, however, it is convenient
to use another dimensionless parameter $\bar{\delta}$, defined as
\begin{equation}
\bar{\delta}=\sqrt{1-\frac{m^{2}}{a^{2(d-3)}}
\frac{(d-1)^{d-1}}{(d-3)^{d-3}}} \ .
\end{equation}
It is clear from proposition 1 that the near extreme
SdS black hole corresponds to the limit
\begin{equation}
0<\bar{\delta}\ll1\label{limit2} \ .
\end{equation}
The limits (\ref{limit1}) and (\ref{limit2}) are
equivalent. Furthermore, it can be shown that they go to zero at the
same rate:
\begin{equation}
\bar{\delta}=\frac{d-1}{2}\delta'+O(\delta'^{2}) \ .
\end{equation}

In the Schwarzschild-de Sitter scenario, the function $h(r)$ has
one local maximum $r_0$ in the interval $[r_{+},r_{c}]$. This point 
can be expressed in terms of the parameters $m$ and $a$ as
\begin{equation}
r_{0}=\left[(d-3)ma^{2}\right]^{1/(d-1)} \ .
\end{equation}
Near the extreme limit, the function $h(r)$
can be approximated by its Taylor expansion up to the second order
in $\delta'$ around the local maximum $r=r_{0}$:
\begin{equation}
h(r)=h(r_{0})+\frac{a^{2}}{2}\left.\frac{d^{2}h(r)}{dr^{2}}
\right|_{r=r_{0}}\left(\frac{r-r_{0}}{a} \right)^{2} 
+ O\left(\delta'^{3}\right) \ .
\label{h-expan}
\end{equation}
To lowest order $\delta'$ and $\bar{\delta}$ are proportional, and 
the expression (\ref{h-expan}) can be written  as
\begin{equation}
h(r)=\frac{d-1}{a^{2}}\left(r_{c}^{ap}-r\right)\left(r-r_{+}^{ap}\right)
+ O\left(\bar{\delta}^{3}\right) \ ,
\label{h-quad}
\end{equation}
where the constants $r_{+}^{ap}$ and $r_{-}^{ap}$ are approximations
of the event and cosmological horizons, given by
\begin{eqnarray}
r_{c}^{ap} & = &
r_{0} +
a\sqrt{\frac{1-\left(1-\bar{\delta}^{2}\right)^{\frac{1}{d-1}}}{d-1}}
\nonumber \\
& = & a\left(\sqrt{\frac{d-3}{d-1}}+\frac{\bar{\delta}}{d-1}\right) +
O\left(\bar{\delta}^{2}\right) \ ,
\end{eqnarray}
\begin{eqnarray}
r_{+}^{ap} & = &r_{0} -
a\sqrt{\frac{1-\left(1-\bar{\delta}^{2}\right)^{\frac{1}{d-1}}}{d-1}}
\nonumber \\
& = & a\left(\sqrt{\frac{d-3}{d-1}}-\frac{\bar{\delta}}{d-1}\right)
+ O\left(\bar{\delta}^{2}\right) \ . 
\end{eqnarray}

The next step is to calculate the tortoise radial function $x(r)$,
defined at the block $T_{+}$ in the usual way,
\begin{eqnarray}
x(r) & = & \int\frac{dr}{h(r)} \nonumber \\
& = & -\frac{1}{2\kappa^{ap}_{c}}\ln\left(r_{c}^{ap}-r\right) +
\frac{1}{2\kappa^{ap}_{+}}\ln\left(r-r_{+}^{ap}\right) \nonumber \\
& &  + O\left(\bar{\delta}^{3}\right) \ , 
\label{x-SdS}
\end{eqnarray}
where the constants $\kappa^{ap}_{c}$ and $\kappa^{ap}_{+}$ are 
\begin{equation}
\kappa^{ap}_{c} = \kappa^{ap}_{+} =
\frac{d-1}{2a^{2}}\left(r_{c}^{ap}-r_{+}^{ap}\right) =
\frac{\bar{\delta}}{a} + O\left(\bar{\delta}^{2}\right) \ .
\end{equation}
Observe that these constants are approximations to the surface gravities
of the event and cosmological horizons, which tend to zero  in the
extreme limit. 

The key point is that, in this limit, the function
$x(r)$ can be analytically inverted. With the expression
for $r(x)$, the function  $h(x) \equiv h\left(r(x)\right)$ can be explicitly
calculated 
\begin{equation}
h(x) = \frac{\bar{\delta}^{2}}{d-1} \frac{1}{\cosh^{2} ( \kappa^{ap}_{+} x )}
+ O\left(\bar{\delta}^{3}\right) \ .
\label{hx-SdS}
\end{equation} 
All the constants in Eq. (\ref{hx-SdS})  can be obtained from the
parameters $m$ and $\Lambda$.   

The approach used here to treat the near extreme Schwarzschild-de
Sitter black hole can be generalized. In the next section we will
see that   $h(x)$  in the expression (\ref{hx-SdS}) have the same form
in a broader class of near extreme spacetimes.

\section{More General Setting}

The fact that the function $h(r)$ for the $d$-dimensional
Schwarzschild-de Sitter black hole can be written in the form
(\ref{h-quad}) is not a particularity of this specific geometry. Basically, all
we have used is the following:  

\begin{itemize}
\item The function $h(r)$ has at least two positive real roots $r_{1}$
and $r_{2}$ ($r_{1}<r_{2}$). If $r_{1}$
and $r_{2}$ are consecutive roots, we are interested in the submanifold
given by the block
\begin{equation}
T_{1}=\left\{
(t,r,\theta_{1},\cdots,\theta_{d-2}), \, \, \, r_{1}<r<r_{2}\right\} \  .
\end{equation}

\item The function $h(r)$ is, at least, a $C^{3}$ function at the interval
$[r_{1},r_{2}]$.

\item The points $r_{1}$ and $r_{2}$ are simple roots of $h(r)$.

\item There is a near extreme limit, a region of the parameter space
where the horizons are arbitrarily close, such that
\begin{equation}
0 < \frac{r_{2} - r_{1}}{r_{1}} \ll 1 \ .
\end{equation}
\end{itemize}

Since $h(r)$ is continuous, there is a maximum or
minimum point $r_{0} \in ]r_{1},r_{2}[$. But in general  the analytical
determination of $r_{0}$ in terms of the parameters of the metric is not
easy. It is convenient then to consider $r_{1}$, $r_{2}$, and
$\kappa_{1}$ as the fundamental parameters of the spacetime, where
$\kappa_{1}$ is the surface gravity at the horizon $r=r_{1}$.

In terms of these parameters, the  function $h(r)$ can be approximated, in
the near extreme limit, as 
\begin{equation}
h(r)=\frac{2 \kappa_{1}}{r_{2} - r_{1}}
(r_{2} - r)(r - r_{1}) + O\left(\delta^{3}\right) \ ,
\label{h-quad2}
\end{equation}
with $\delta = (r_{2}-r_{1})/r_{1}$. The tortoise function $x(r)$,
whose domain is the interval $]r_{1},r_{2}[$,
can be easily calculated from the expression (\ref{h-quad2}). From its
inverse  $r(x)$ it is straightforward to obtain $h(x)$:
\begin{equation}
h(x)=\frac{(r_{2} - r_{1}) \kappa_{1}}{2 \cosh^{2} (\kappa_{1} x )}
+ O\left(\delta^{3}\right) \ .
\end{equation}

One possible generalization of the $d$-dimensional Schwarzschild-de
Sitter geometry is  the metric in the form (\ref{metric}) with the function
$h(r)$ given by
\begin{equation}
h(r)=1 - \frac{2m}{r^{d-3}} + \frac{q^{2}}{r^{2d-6}} - 
     \frac{\Lambda r^{2}}{3} \ .
\end{equation}
This is the $d$-dimensional Reissner-Nordstr\"{o}m-de Sitter 
metric \cite{Thangherlini-63}, which describes a charged black hole
asymptotically de Sitter. The integration constants  $m$ and $q$ are
proportional to the  black hole mass  and electric charge.  

\newpage

The parameter space for the RNdS metric is much more complex than the
SdS case. It can be shown that the function $h(r)$, for arbitrary $d$,
has at most three positive real roots (plus negative roots). And if
the number of positive roots is three, they are simple. These roots
are the Cauchy ($r_{-}$),  event ($r_{+}$), and cosmological ($r_{c}$)
horizons, with $0<r_{-}<r_{+}<r_{c}$. The function $h(r)$ is smooth in
both  intervals $]r_{-},r_{+}[$ and $]r_{+},r_{c}[$, and there are
regions of the parameter space in which the intervals collapse to
points.   

The conditions shown in the beginning of this section therefore apply
to the RNdS metric, and we have two possible near extreme situations
for the RNdS metric, where $r_{-}\approx r_{+}$ and $r_{+}\approx
r_{c}$. Although the work developed in this section applies for both
cases, we are mainly interested in the second one. We will therefore
specify the work to the near extreme $T_{+}$ block of the
Schwarzschild-de Sitter and Reissner-Nordstr\"{o}m-de Sitter metric.

\section{Effective Potential and Quasinormal Modes}

We introduce  in the $d$-dimensional SdS  or RNdS spacetime
a real scalar field $\Phi$,  with mass $\mu \geq 0$,  described  by
the equation  
\begin{equation}
\left( \Box - \mu^2 \right) \Phi=0 \ .
\end{equation}
Expanding the field in (hyper)spherical harmonics, in the form
\begin{equation}
\Phi=\sum_{\ell \, m} r^{-(d-2)/2}\psi_{\ell}(t,r)
\textrm{Y}_{\ell\, m}(\{\theta_{i}\}) \ ,
\label{Ansazs_field}
\end{equation}
we get a decoupled wave equation for each value of $\ell$. In terms of
the coordinates $t$ and $x$, this equation reads
\begin{equation}
-\frac{\partial^{2}\psi_{\ell}}{\partial
 t^{2}}+\frac{\partial^{2}\psi_{\ell}}{\partial
 x^{2}}=V(x)\psi_{\ell} \ .
\label{wave_equation}
\end{equation}

The effective potential $V(x) \equiv V(r(x))$ is obtained
using the function $r(x)$ and the effective potential in terms of
the radial coordinate $r$:
\begin{equation}
V(r)=h(r)\Omega(r) \ ,
\end{equation}
with the function $\Omega(r)$ given by 
\begin{eqnarray}
\Omega(r) & = & \mu^2 +
\frac{\ell(\ell+d-3)}{r^{2}}+\frac{d-2}{2r}h'(r) \nonumber \\
& & + \frac{(d-2)(d-4)}{4r^{2}} h(r) \ .
\end{eqnarray}

Expanding the function $V(r)$ around the point $r = (r_{+} +
r_{c})/2$, we have, for $\ell>0$ or $\mu>0$: 
\begin{equation}
V(r)  =  \left[ \frac{\ell(\ell+d-3)}{r_{+}^{2}} + \mu^{2} \right]
\frac{2 \kappa_{+}  (r_{c} - r)(r - r_{+})}{r_{c} - r_{+}}
+ O(\delta) \ .
\label{V-r-general}
\end{equation}
The expression (\ref{V-r-general}) for $V(r)$ 
is not useful when $\mu=\ell=0$ because in this case $V(r) = 0 +
O(\delta)$. Assuming $\ell>0$ or $\mu>0$, we calculate the
function $V(x)$ as   
\begin{equation}
V(x)=\frac{V_{0}}{\cosh^{2}\left(\kappa_{+}x\right)}
+ O(\delta )
\ ,
\label{V-general}
\end{equation}
where the constant $V_{0}$ is given by
\begin{equation}
V_{0}= \left[ \frac{\ell(\ell+d-3)}{r_{+}^{2}} + \mu^{2} \right]
\frac{(r_{c} - r_{+})\kappa_{+}}{2} \ .
\label{V0-general}
\end{equation}

The effective potential $V(x)$ in the expression (\ref{V-general}) is the
P\"{o}schl-Teller potential \cite{Poschl-33}, which has been extensively
studied. In particular, attention has been dedicated to the
quasinormal modes associated \cite{Ferrari-84,Beyer-99}. One way to
introduce the problem \cite{Kokkotas-99,Nollert-92} is Laplace
transforming the wave equation (\ref{wave_equation}), so the problem
can be put as a Cauchy initial value problem. It is found that there
is a discrete set of possible values to $s$ such that the function
$\hat{\psi}_{\ell}$, the Laplace transformed field, satisfies both
boundary conditions:  
\begin{equation}
\lim_{x \rightarrow -\infty}\hat{\psi}_{\ell} \, e^{sx}=1 \ ,
\end{equation}
\begin{equation}
\lim_{x \rightarrow +\infty}\hat{\psi}_{\ell} \, e^{-sx}=1 \ .
\end{equation}
By making the formal replacement $s=i\omega$, we have the usual quasinormal
mode boundary conditions. The frequencies $\omega$ (or $s$) are
called quasinormal frequencies.

Using the result (\ref{V0-general}), we find that for both
Schwarzschild-de Sitter and Reissner-Nordstr\"{o}m-de Sitter cases we
have
\begin{widetext}
\begin{equation}
\frac{\omega}{\kappa_{+}} =  \sqrt{\left[ \frac{\ell(\ell+d-3)}{r_{+}^{2}} +
\mu^{2} \right] \frac{r_{c} - r_{+}}{2\kappa_{+}} - \frac{1}{4}}
-i \left(n+\frac{1}{2}\right)  \ . 
\end{equation}
\end{widetext}
For the SdS geometry, $V_{0}$ can be written in terms of  $m$ and
$\Lambda$. The real and imaginary parts of the quasinormal frequencies
in this case are  
\begin{eqnarray}
\textrm{Re}(\omega) & = & \left[
\frac{\Lambda}{3}-\frac{m^{2}\Lambda^{d-2}}{3^{d-2}} 
\frac{(d-1)^{d-1}}{(d-3)^{d-3}} \right]^{\frac{1}{2}} \nonumber \\
& & \left[ 
\frac{\ell(\ell + d - 3)}{d-3} + \frac{3\mu^2}{(d-1)\Lambda} 
    - \frac{1}{4} \right]^{\frac{1}{2}} \ ,
\end{eqnarray}
\begin{equation}
\textrm{Im}(\omega)  =  - \left(n+\frac{1}{2}\right) 
\left[ \frac{\Lambda}{3}-\frac{m^{2}\Lambda^{d-2}}{3^{d-2}}
\frac{(d-1)^{d-1}}{(d-3)^{d-3}} \right]^{\frac{1}{2}} \ , 
\end{equation}
with $n \in \{0,1,\ldots\}$ labelling the modes.

\section{Conclusions}

We have studied  a scalar field  outside the
event horizon of spherical $d$-dimensional black holes with near
extreme  cosmological constant. Its dynamics is  determined by a
P\"{o}schl-Teller effective potential, which allows us to calculate
analytic expressions for the quasinormal frequencies. In the
Schwarzschild-de Sitter case, the parameter space  can be
precisely characterized. As a consequence, the quasinormal modes can
be written in terms of the  parameters $m$ and $\Lambda$ of the
metric.    

Our results generalize the previous conclusions obtained in
\cite{Cardoso-03}.  In particular, we see that the real part of the
quasinormal frequencies does not depend on the mode ($n$), and that
the relaxation time of the field is independent of its mass. We also
demonstrate that the addition of charge to the black hole or mass to
the scalar field does not  alter the basic characteristics of the
field dynamics in the near extreme regime. 
  
Since we are imposing spherical symmetry, it is not too surprising that the
field evolution is qualitatively independent of the dimension. However, it is
interesting to note that the explicit expressions of the frequencies are very
similar for any value of $d$. It would be instructive to see if this
happens in non-spherical geometries. 


\begin{acknowledgments}
The author is grateful to E. Abdalla for very useful discussions and
encouragement. This work was supported by \emph{Funda\c{c}\~{a}o de Amparo
\`{a} Pesquisa do Estado de S\~{a}o Paulo (FAPESP)}, Brazil.
\end{acknowledgments}


%
%
\end{document}